\documentclass[aps,prb,superscriptaddress,twocolumn,floatfix]{revtex4}

\usepackage{graphicx,color}
\usepackage{dcolumn}
\usepackage{bm}
\usepackage{stmaryrd}
\usepackage{latexsym}
\usepackage{amssymb}
\usepackage{amsfonts}
\usepackage{amsmath}

\begin{document}

\title{Selective doping Barlowite for quantum spin liquid: a first-principles study}

\author{Zheng Liu}
\affiliation{Institute for Advanced Study, Tsinghua University, Beijing 100084, China}
\affiliation{Department of Materials Science and Engineering, University of Utah, Salt Lake City, UT 84112, USA}

\author{Xiaolong Zou}
\affiliation{Department of Materials Science and Nanoengineering, Rice University, Houston, TX 77005, USA}

\author{Jia-Wei Mei}
\affiliation{Perimeter Institute for Theoretical Physics, Waterloo, Ontario, N2L 2Y5 Canada}

\author{Feng Liu}
\affiliation{Department of Materials Science and Engineering, University of Utah, Salt Lake City, UT 84112, USA}
\affiliation{Collaborative Innovation Center of Quantum Matter, Beijing 100084, China}

\date{\today}


\begin{abstract}
Barlowite $Cu_4(OH)_6FBr$ is a newly found mineral containing $Cu^{2+}$  kagome planes. Despite similarities  in many aspects to Herbertsmithite $Cu_3Zn(OH)_6Cl_2$, the well-known quantum spin liquid (QSL) candidate, intrinsic Barlowite turns out not to be a QSL, possibly due to the presence of  $Cu^{2+}$ ions in between kagome planes that induce interkagome magnetic interaction [PRL, 113, 227203 (2014)]. Using first-principles calculation, we systematically study the feasibility of selective substitution of the interkagome Cu ions with isovalent nonmagnetic ions. Unlike previous speculation of using larger dopants, such as $Cd^{2+}$ and $Ca^{2+}$, we identify the most ideal stoichiometric doping elements to be Mg and Zn in forming $Cu_3Mg(OH)_6FBr$ and $Cu_3Zn(OH)_6FBr$ with the highest site selectivity and smallest lattice distortion.  The equilibirium anti-site disorder in Mg/Zn- doped Barlowite is estimated to be one order of magnitude lower than that in Herbertsmithite. The single-electron band structure and orbital component analysis show that the proposed selective doping effectively mitigates the difference between Barlowite and Herbertsmithite.
\end{abstract}

\maketitle

Quantum spin liquid (QSL) represents a new state of matter characterized by long-range entanglement, beyond the conventional symmetry-breaking paradigm \cite{XiaogangWen}. Realizing QSL in real-world materials has been a long-sought goal for decades \cite{Anderson73RVB, Nature10Balentsqsl, Science08PALeeQSL}. The most promising candidate so far is Herbertsmithite $Cu_3Zn(OH)_6Cl_2$, which realizes the $S = 1/2$ antiferromagnetic (AFM) Heisenberg model on the 2D kagome lattice \cite{Nocera05Herbert}. Extensive theoretical studies have suggested that this model is likely to achieve a QSL ground-state, despite close in energy with other competing phases \cite{Sachdev92kagomedisorder,PRB97disorderkagome, Matthew07kagomeliquid,RanWen07Kagome, Lu11Kagome, Yan11DMRGkagome,PRB13Yasir1,PRB13Yasir2}.  Experiments on Herbertsmithite have also shown QSL-like features, such as the absence of any observed magnetic order down to 50 mK \cite{YLee07Neutron,Mendel07HerbertUSR} and an unusual continuum of spin excitations \cite{YLee12Neutron}.  However, the inevitable Cu/Zn antisite disorder makes the interpretation of experimental data difficult  \cite{Harrison08NMRHerbert}. It remains an open debate whether these defects obscure the intrinsic signals, such as a tiny spin gap that is crucial for the classification of the ground state \cite{Mendels04Reivew}.

Very recently, Barlowite $Cu_4(OH)_6FBr$ as a new kagome compound was discovered \cite{Mine10Barlow}. Its structure closely resembles Herbertsmithite, whereas the Cu/Zn antisite disorder is automatically avoided. Therefore, studies on this new material are expected to shed fresh light on understanding the kagome physics and QSL phase. Interestingly, Barlowite is diagnosed with a Curie-Weiss constant $\theta_{CW}$=-136K close to Herbertsmithite, yet it  undergoes a spin-ordering phase transition at 15K\cite{Han14Barlow}. The low-temperature magnetic properties were further investigated by combining first-principles calculation with experiments \cite{Arxiv14Valenti}. Since the main structural difference between these two materials is the cations occupying the interkagome sites, i.e. $Cu^{2+}$ and $Zn^{2+} $ in Barlowite and Herbertsmithite, respectively, it is suggested that substituting the interkagome sites with nonmagnetic ions should tune Barlowite into the same phase as Herbertsmithite. Specifically, relatively larger elements, such as Sn and Cd, are speculated as possible candidates for substitution based on the simple argument of lattice spacing of the interkagome sites \cite{Han14Barlow}.

\begin{figure}[ht]
\includegraphics[width=0.4\textwidth]{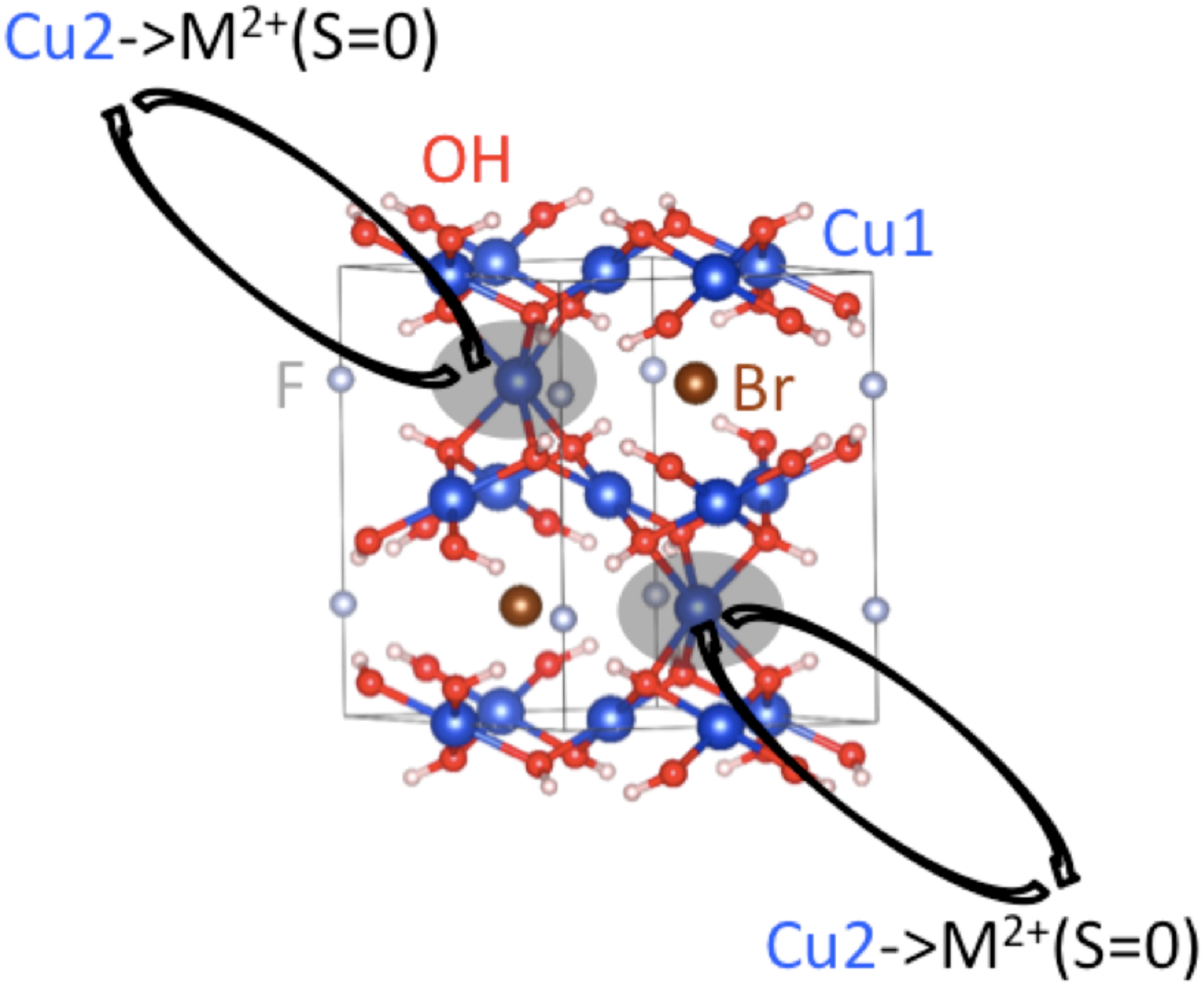}
\caption{\label{fig:structure}Unit cell and atomic structure of Barlowite, and the schematic illustration of the proposed selective doping.}
\end{figure}

In this Letter, we identify the most promising candidates for realizing the selective doping to form stoichiometric doped Barlowite, based on density functional theory (DFT) calculations \cite{RMP89DFT}. We systematically calculate the doping energies and analyze the doping selectivity of a series of nonmagnetic group 2 and 12 elements. Unlike the previous speculation \cite{Han14Barlow}, larger dopants are found to have lower site selectivity and tend to distort the kagome plane more than smaller dopants. Most importantly, we identify Mg and Zn to be the most ideal choices of dopants to form stoichiometric $Cu_3Mg(OH)_6FBr$ and $Cu_3Zn(OH)_6FBr$ compounds in the Barlowite family, with the highest site selectivity in substituting the interkagome Cu ions and the least lattice distortion in kagome planes. Statistical analysis shows that the equilibrium distribution of Mg/Zn in the Mg/Zn-doped Barlowite at the typical growth temperature exhibits a level of antisite disorder significantly lower than that in Herbertsmithite. Also, single-electron band structures of intrinsic and doped Barlowite are calculated, and discussed in comparison with Herbertsmithite.

Figure 1 shows the atomic structure of Barlowite. Similar to Herbertsmithite, it contains $Cu^{2+}$  kagome planes connected by hydroxyls. The difference lies in the interkagome site: in Barlowite, there are additional $Cu^{2+}$ ions between the kagome planes, which act as additional spin 1/2 centers and mediate interkagome spin exchange. Therefore, to clarify the different magnetic ground states between Barlowite and Herbertsmithite, one way is to remove these out-of-plane spins by selective doping. The chosen dopants should be spin zero and isovalent.  Using these two criteria, we have considered elements of $Mg$, $Ca$, $Sr$, $Ba$ from group 2 and $Zn$, $Cd$ from group 12. The ionic radius ($r_i$)and standard chemical potential in aqueous solution($\mu^0_i$) of these ions are listed in Tab. I. Note that $Mg^{2+}$ and $Zn^{2+}$ are close in radius to $Cu^{2+}$, while the other ions are larger. In addition, all these dopants under consideration have a smaller $\mu^0_i$  than $Cu^{2+}$, indicating a stronger tendency to stay in the solution.

\begin{table}[ht]
\caption{Ionic radius ($r_i$) and  standard chemical potential in aqueous solution ($\mu^0_i$) of Cu and the dopants under consideration}
\begin{ruledtabular}
\begin{tabular}{ccc}
 & $r_i$ (pm) \cite{IonicRadius} &$\mu^0_i$ (eV) \cite{muion}\\
  \hline
 Cu$^{2+}$ & 73 & 0.68 \\
 \hline
 Mg$^{2+}$ & 72 & -4.73\\
 Ca$^{2+}$ & 100 & -5.73 \\
 Sr $^{2+}$& 118 & -5.78 \\
 Ba$^{2+}$ & 135 & -5.81\\
 \hline
 Zn$^{2+}$ & 74 & -1.52\\
 Cd$^{2+}$ & 95 & -0.80\\
\end{tabular}
\end{ruledtabular}
\end{table}

The doping process is expected to take place by adding dopant ions in solution during the hydrothermal growth. The net reaction equation is written as : 
\begin{eqnarray}
\begin{split}
(1-x_d)Cu_4(OH)_6FBr+4x_dM^{2+}+6x_dOH^-\\
+x_dF^-+x_dBr^- \rightarrow [Cu_{1-x_d}M_{x_d}]_4(OH)_6FBr
\end{split}
\end{eqnarray}
in which $M^{2+}$ denotes the dopants. 

Equation (1) can be considered as a combination of  two subreactions:
\begin{eqnarray}
\begin{split}
Cu_4(OH)_6FBr+4x_dM^{2+} \leftrightharpoons \\
[Cu_{1-x_d}M_{x_d}]_4(OH)_6FBr +4x_dCu^{2+} 
\end{split}
\end{eqnarray}
\begin{eqnarray}
\begin{split}
4x_dCu^{2+}+6x_dOH^-+x_dF^-+x_dBr^-\\
\rightarrow x_dCu_4(OH)_6FBr
\end{split}
\end{eqnarray}
Equation (2) describes the simple substitution process. Then, after the Cu ion is exchanged into the solution, the overall free energy of the system can be lowered by forming more deposits of Barlowite [Eq.(3)]. The latter provides the thermodynamical driving force for the doping process to proceed.  Equation (3) is actually nothing but the growth of undoped Barlowite as reported in previous experiments \cite{Han14Barlow,Arxiv14Valenti}. Hence, we will focus on evaluating the experimental feasibility of Eq.(2) only. 

The central physical quantity we are going to calculate is the standard doping energy $E_{d}$ as defined by the total energy difference per substitution.  According to Eq.(2), $E_d$ consists of two parts: $E_d=\Delta E_B+\Delta \mu_{i}^0$, where $\Delta E_B=(E_{[Cu_{1-x_d}M_{x_d}]_4(OH)_6FBr}-E_{Cu_4(OH)_6FBr})/4x_d$ is the energy change of Barlowite after doping and $\Delta \mu_i^0=\mu^0_{Cu^{2+}}-\mu^0_{M^{2+}}$ is the standard chemical potential difference between the two ions (Tab. I). There are two inequivalent doping sites: Cu1 is in the kagome plane; Cu2 is between the kagome planes. We use $E_{d1}$ and $E_{d2}$ to differentiate these two types of doping energies. Their difference $\Delta E_d(x_d)=E_{d1}(x_d)-E_{d2}(x_d)$ tells the site preference for dopants, i.e. defines the degree of selective doping.

\begin{figure}[ht]
\includegraphics[width=0.47\textwidth]{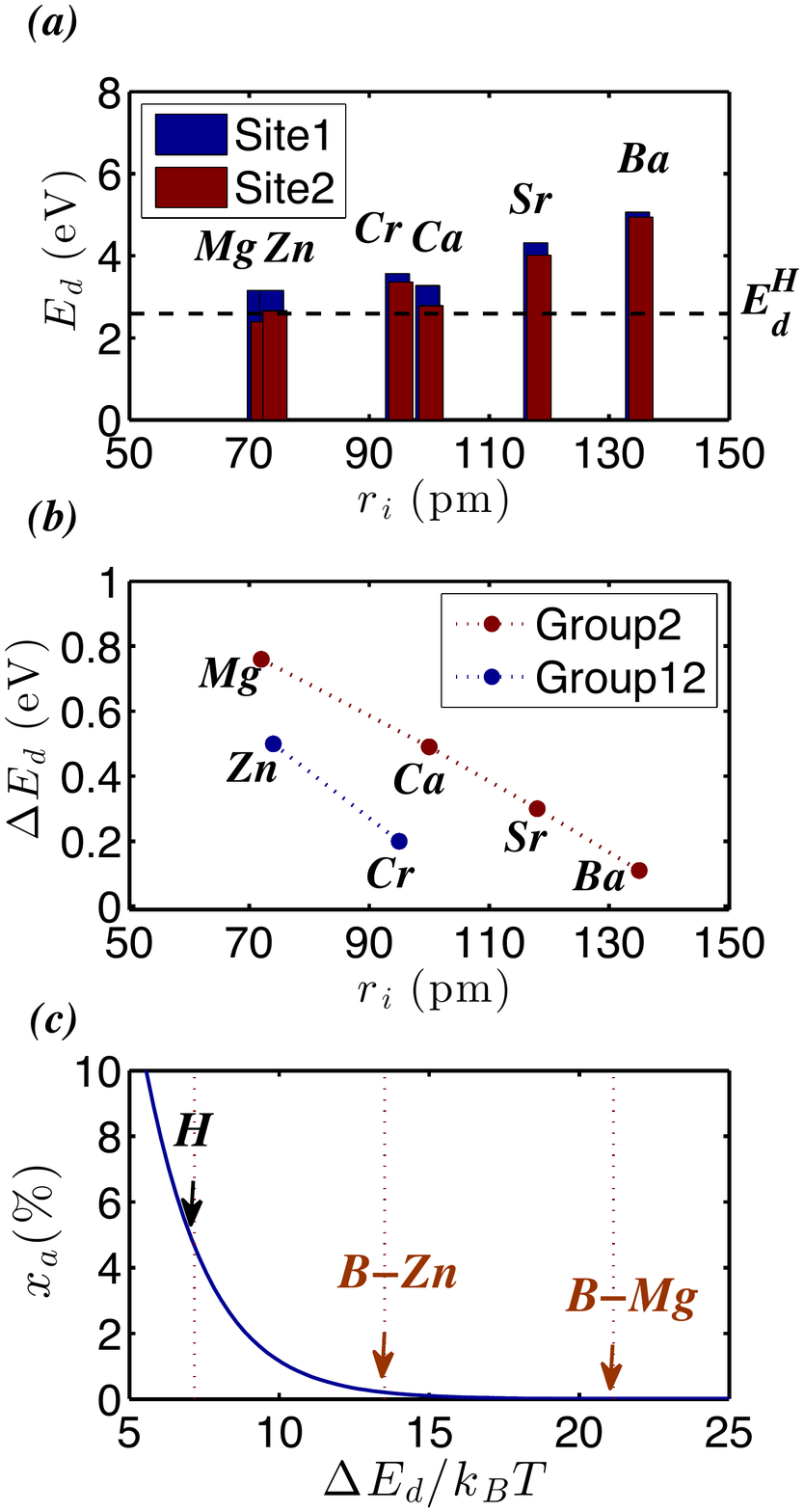}
\caption{\label{fig:Ed} (a) $E_d$ and (b) $\Delta E_d$ of different dopants. $E_d^H$ in (a) corresponds to the standard doping energy for the growth of Herbertsmithite. The dotted line in (b) only serves as a guide to eyes.  (c) Equilibrium anti-site disorder of Herbertsmithite (H)  and Zn/Mg doped Barlowite (B-Zn/Mg) }
\end{figure}

Our calculation on $\Delta E_B$ is carried out using the VASP package \cite{VASP96}, which solves the DFT Hamiltonian self-consistently using the plane wave basis together with the projector augmented wave method \cite{VASPpaw99}. A plane-wave cutoff of 500 eV is enforced.  The self-consistent iterations are converged to 0.1 meV precision of the total energy. We employ the standard generalized gradient approximation (GGA) for the exchange-correlation functional, which is known to satisfactorily describe ionic bonding and cohesive energy even for transition elements \cite{PBE96}.  Barlowite stays in the paramagnetic phase above 15K (1meV) \cite{Han14Barlow,Arxiv14Valenti}. The Dzyaloshinskii-Moriya (DM) interaction arising from spin-orbit coupling is estimated to be of the order of 1 meV \cite{Arxiv14Valenti}. Therefore, a spineless calculation on the structure and total energy is reasonable. As a benchmark, Table 2 summarizes the calculated structural parameters of undoped Barlowite, which agree with the experimental results well. Note that we do not intend to discuss strong correlation effects associated with the Cu 3d orbitals within this methodology.

To simulate doping, we construct a $2\times2\times1$ supercell containing 24 in-plane Cu1 sites and 8 interkagome Cu2 sites, and replace one of the Cu with a dopant. We first fix lattice contants to experimental values, while let the atomic coordinations fully relax until the forces are less than 0.05 eV$/\AA$. The total-energy integration over the Brillouin zone is obtained on a $\Gamma$-centered $2\times2\times 2$ k-mesh.

\begin{table}[ht]
\caption{A comparison of structural parameters between theory and experiment }
\begin{ruledtabular}
\begin{tabular}{cccc}
 Barlowite & Exp. \cite{Han14Barlow} & Exp. \cite{Arxiv14Valenti}& Cal.\\
  \hline
 Lattice a/b ($\AA$)  & 6.68 & 6.80 & 6.73 \\
 Lattice c ($\AA$) & 9.31 & 9.31 & 9.47 \\
 Angle Cu1-O-Cu1 & 117.4$^\circ$ &  117$^\circ$& 117.3$^\circ$ \\
 Angle Cu1-O-Cu2 & 95.8$^\circ$ &  &96.6$^\circ$\\
\end{tabular}
\end{ruledtabular}
\end{table}

Figure 2(a) shows the calculated $E_d$ of different dopants occupying the Cu1 and Cu2 sites.  As a reference, we have also calculated the standard doping energy for the growth of Herbertsmithite [$E_d^H$ in Fig. 2(a)] as described by the following equation: $[Cu_{3+x_d}Zn_{1-x_d}](OH)_6Cl_2+x_dZn^{2+} \leftrightharpoons Cu_3Zn(OH)_6Cl_2+x_dCu^{2+}$. The positive doping energy acts as a reaction barrier, which limits the kinetics of Eq.(1). The value of $E_{d2}$ for Zn is found almost the same as $E_{d}^H$, again reflecting the similarities between Barlowite and Herbertsmithite.  $E_d$ typically increases with the ionic radius as a size effect: Mg is even easier to substitute Cu than Zn, while larger dopants are more difficult. 

The kagome-site doping energy ($E_{d1}$) is always higher than the interkagome-site doping energy ($E_{d2}$), indicating that the latter is the preferred site for doping. The preference for dopants to occupy the interkagome site provides exactly the type of doping selectivity we need. It can be understood by noticing the larger space around the Cu2 site than the Cu1 site (Fig.1). In Fig. 2(b), we plot $\Delta E_{d}=E_{d1}-E_{d2}$ as a function of ion radius,  showing directly the site preference in relation to ionic radius. Within each group, $\Delta E_{d}$ decreases monotonously. The elements from two groups are however not on the same line. This result  is reasonable considering that the outermost shells of group 2 and 12 ions are p-electrons and d-electrons, respectively. Thus, they will exhibit different interfacial bonding energy with the surrounding. One important outcome revealed in Fig. 2(d) is that $\Delta E_{d}$ decreases for larger dopants. Therefore, large dopants actually have lower site preference, hence are more difficult to achieve stoichiometric doping. This invalidates the previous speculation \cite{Han14Barlow}.

One more problem for the large dopants is identified after fully relaxing the lattice volume and geometry. While interkagome doping maintains the original lattice symmetry, the in-plane doping distorts the lattice from hexagonal to triclinic, breaking the perfect kagome plane and lowering the  in-plane doping energy.  Consequently, $\Delta E_d$ becomes smaller. This effect becomes very significant for larger dopants. For example, for $Cd^{2+}$  and $Ca^{2+}$,  $\Delta E_d$ decreases from 0.20 and 0.50 eV to -0.12 and 0.11 eV, respectively. It means that upon doping, a large fraction of dopants will substitute the in-plane sites, which in turn distorts the lattice and hinders stoichiometric selective doping. In contrast,  for $Mg^{2+}$  and $Zn^{2+}$, which have similar radius to $Cu^{2+}$,  this problem does not occur: $\Delta E_d$ decrease from 0.76 and 0.50 eV to 0.72 and 0.46 eV, respectively, which are still sufficiently large to suppress  in-plane doping. Therefore, we conclude that for our purpose $Mg^{2+}$  and $Zn^{2+}$ are the most ideal dopants.

To further examine whether the selective doping of Mg/Zn at the interkagome sites can sustain up to the stoichiometric limit to form $Cu_3Mg(OH)_6FBr$ and $Cu_3Zn(OH)_6FBr$ compounds, we proceed by progressively increasing the amount of interkagome dopants. $E_d$ is found to be nearly independent of the doping concentration ($x_d$). This property indicates that the interaction between the dopants is weak, which is important for reaching the the stoichiometric limit. Otherwise, dopants may form clusters or hinder further doping process. We have also checked that the lattice and Cu kagome planes in  $Cu_3Mg(OH)_6FBr$ and $Cu_3Zn(OH)_6FBr$ remain stable under structural relaxation.

It is worth making some comparison to Herbertsmithite within the present methodology. The first critical issue is the degree of equilibrium anti-site disorder in the two systems.
The effect of Cu/Zn anti-site disorder in Herbertsmithite has been a long-lasting debate \cite{Mendels04Reivew}.  Such disorder is inevitable in doped Barlowite as well, because it leads to an increase of the configuration entropy. Under constant temperature and volume, the equilibrium is reached by minimizing the free energy $F(x_a)=E(x_a)-TS(x_a)$ with respect to the anti-site concentration $x_a$. Without loss of generality, it is convenient to set E(0)=S(0)=0 for the stoichiometric systems. Accordingly, when $x_a$ pairs of Mg(Zn) and Cu per empirical formula switch sites, the energy increase is simply $E(x_a)=x_a\Delta E_d$, given that the interaction between the dopants is weak. The entropy increase per empirical formula can be analytically derived as $S(x_a)=k_B\ln\Omega_a$, where $\Omega_a$ is the anti-site configuration number. With the aid of the Sterling's approximation, the final form of entropy is $S(x_a)=-k_B\ln[(3-x_a)^{3-x_a}(1-x_a)^{1-x_a}x_a^{2x_a}]$. The minimal point of $F(x_a)$ is calculated numerically, which can be expressed as a function of $\Delta E_d/k_BT$ [Fig. 2(c)]. By using the calculated values of $\Delta E_d$ (Mg: 0.72 eV; Zn: 0.46 eV) and the experimental growth temperature (T=393K) \cite{Han14Barlow}, the anti-site disorder in Mg/Zn doped Barlowite is predicted to be below $0.1\%$. For comparison, we have also calculated $\Delta E_d=0.30 eV$ for Herbertsmithite and with T=483K \cite{Nocera05Herbert}, the equilibrium disorder is calculated  to be $5\%$, comparable to the experimental estimation \cite{Mendels04Reivew}. Therefore, the degree of anti-site disorder in the doped Barlowite is expected to be at least one order of magnitude lower than that in Herbertsmithite, owing to a higher $\Delta E_d$ as well as a lower growth temperature. This difference can be significant to help clarify the effects of disorder on the QSL phase.

Secondly, we do a comparison of single-electron band structure. Figure 3 shows the band structures of Hebertsmithite, undoped and doped ($x_d=1$) Barlowite marked with orbital compositions. Despite the absence of strong-correlation effects, the DFT single-electron band structure  properly describes single-electron hopping processes, which serve as the guide to the AFM superexchange.  For Herbertsmithite, there is a set of bands around the Fermi level (between -0.5 eV and 0.75 eV), gapped from the underlying valence bands \cite{DFT13Herbert} . These bands primarily arise from Cu  [blue cross in Fig.3(a)] and the adjacent O (not shown), exhibiting the typical features of NN hopping on a 2D kagome lattice \cite{Zheng14CPB}. For Barlowite [Fig.3(b)], around 0.5 eV the band dispersion is similar to that in Herbersmithite with the band composition primarily from the in-plane Cu, indicating similar hopping amplitude within the kagome planes. This is in agreement with the experimental fact that the Curie-Weiss constants for Herbersmithite and Barlowite are close \cite{Han14Barlow}. However, around the Fermi level, interkagome Cu not only contributes extra bands, but also strongly mix with the Cu1 bands. This result suggests considerable coupling between Cu1 and Cu2, as pointed out by previous studies \cite{Han14Barlow,Arxiv14Valenti}.  The effect of replacing Cu2 with Zn or Mg is remarkable [Figs.3(c) and (d)]. After doping, the complexities of interkagome coupling are removed. Both Zn and Mg states are far from the Fermi level, leaving clean Cu1 bands around the Fermi level. The overall band dispersion also becomes closer to Herbertsmithite.   The energy states below the Fermi level contain contribution from the halogen atoms, i.e. Cl and Br in Herbertsmithite and Barlowite, respectively.   We note that both $Cl^-$ and $Br^-$ are spin zero and far away from the superexchange path between Cu1 ions. Therefore, these orbitals do not play an important role in the magnetic properties of the materials.

\begin{figure*}[ht]
\includegraphics[width=0.9\textwidth]{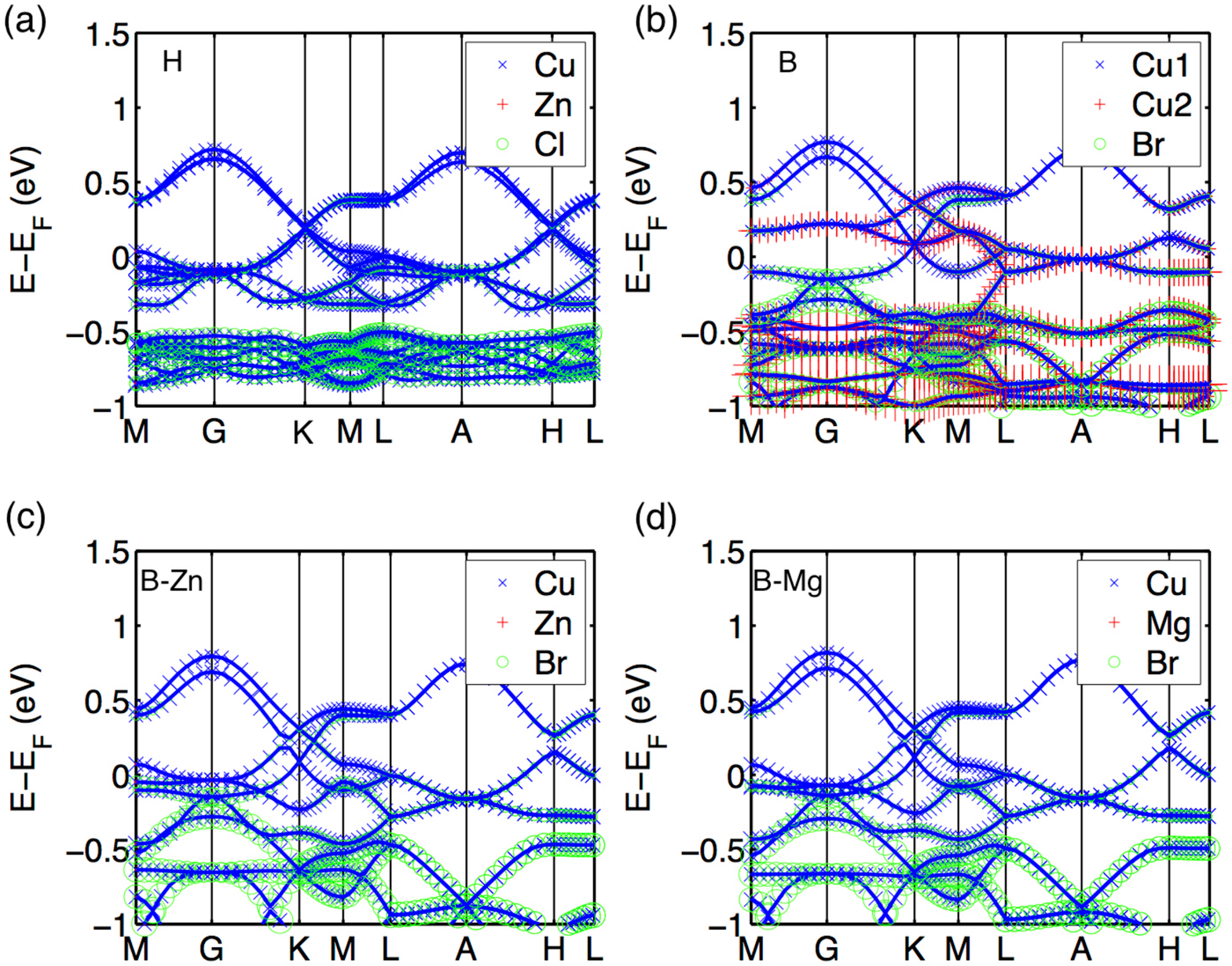}
\caption{\label{fig:band} Orbital resolved single-electron band structure of Herbertsmithite (H), Barlowite (B) and Zn/Mg doped Barlowite (B-Zn/Mg). The marker size reflects the weight of atomic composition. }
\end{figure*}

In conclusion, based on the DFT calculations, we identify $Cu_3Mg(OH)_6FBr$ and $Cu_3Zn(OH)_6FBr$ as the most promising targets to realize the stoichiometric doped Barlowite. The distinct advantages include no lattice distortion, high site selectivity and low anti-site disorder. The standard doping energy is comparable to (for Zn) or even lower (for Mg) than that for growing Herbertsmithite. Therefore, these targets may be readily synthesized using similar experimental conditions as used for Herbertsmithite. The remaining open question is how the doped Barlowite behaves magnetically under low temperature: will it be tuned into the same phase as Herbertsmihite or stay as the undoped Barlowite. For either case, the effective doping of this new material as we propose here serves as a useful guide to future experiments in a pursuit to reveal key factors towards QSL.

\title{Selective doping Barlowite for quantum spin liquid: a first-principles study}

\end{document}